\documentclass[12pt,twoside,fleqn]{article}
\usepackage{epsfig}
\usepackage{multirow}
\def\lsim{\raise0.3ex\hbox{$<$\kern-0.75em\raise-1.1ex\hbox{$\sim$}}}
\def\gsim{\raise0.3ex\hbox{$>$\kern-0.75em\raise-1.1ex\hbox{$\sim$}}}
\makeatletter
\@addtoreset{equation}{section}
\makeatother

\renewcommand{\arraystretch}{1.25}
\newcommand{\beqn}{\begin{equation}}
\newcommand{\eqn}{\end{equation}}
\newcommand{\bqa}{\begin{eqnarray}}
\newcommand{\eqa}{\end{eqnarray}}
\newcommand{\bqas}{\begin{eqnarray*}}
\newcommand{\eqas}{\end{eqnarray*}}
\newcommand{\bdm}{\begin{displaymath}}
\newcommand{\edm}{\end{displaymath}}

\newcommand{\Tr}{ {}{\rm Tr}{} }
\newcommand{\tr}{\mbox{Tr~}}
\newcommand{\re}{\mbox{Re~}}
\newcommand{\nn}{\nonumber}
\newcommand{\plaq}{\mbox{\raisebox{-1.25mm}
{\epsfig{file=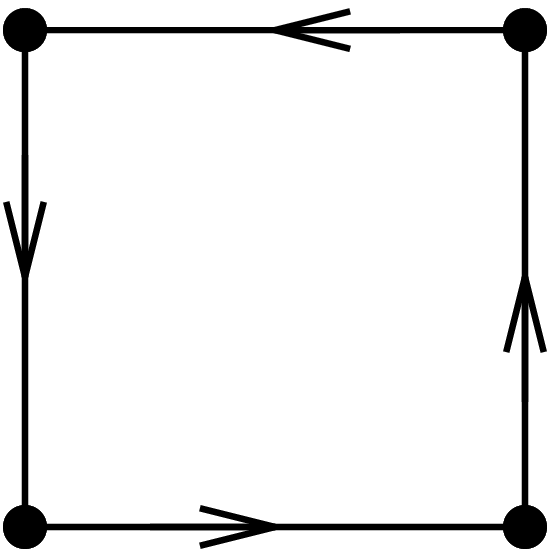,height=5mm
}}~}}
\newcommand{\loOp}{\mbox{\raisebox{-1.25mm}
{\epsfig{file=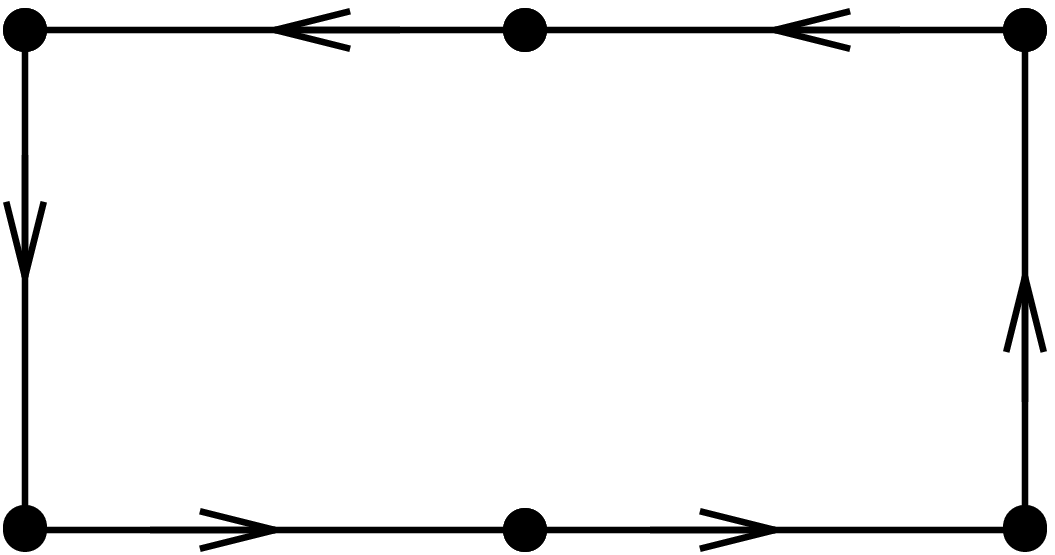,height=5mm
}}~}}
\newcommand{\lOop}{\mbox{\raisebox{-4mm}
{\epsfig{file=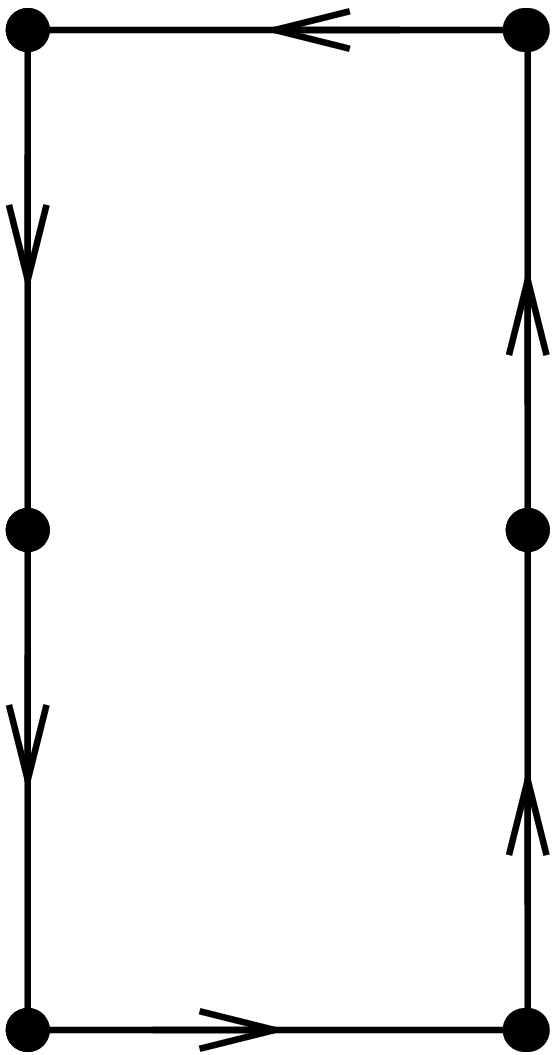,height=10mm
}}~}}
\newcommand{\alink}{\mbox{
\begin{picture}(2.5,.2)
\linethickness{1mm}
\multiput(0,0.1)(2,0){2}{\circle*{0.1}}
\multiput(1.01,0.1)(0,0){1}{\circle{0.2}}
\put(0.8,0.){\link}
\put(-0.2,0.){\linka}
\put(1.1,-.2){\scriptsize \( x \) }
\put(0,-.2){\scriptsize \( y \) }
\put(2.0,-.2){\scriptsize \( y \) }
\end{picture}}}
\newcommand{\alinkfat}{\mbox{
\begin{picture}(3.0,.2)
\linethickness{1mm}
\multiput(0,0.1)(2.4,0){2}{\circle*{0.1}}
\multiput(1.21,0.1)(0,0){1}{\circle{0.2}}
\put(1.2,-1.){\staple}
\put(-0.2,-1.){\staplea}
\put(1.1,-.3){\scriptsize \( x \) }
\put(-.4,-.2){\scriptsize \( y \) }
\put(2.6,-.2){\scriptsize \( y \) }
\end{picture}}}
\newcommand{\blinkba}{\mbox{
\begin{picture}(2.5,2.5)
\thicklines
\multiput(1,-1)(0,-1){2}{\circle*{0.1}}
\multiput(1,0)(0,0){1}{\circle{0.2}}
\multiput(2,0.0)(0,1){3}{\circle*{0.1}}
\multiput(0,-2)(0,0){1}{\circle*{0.1}}
\put(1,-2.0){\vector(-1,0){1}}
\put(1,-1.0){\vector(0,-1){1}}
\put(1,0.0){\vector(0,-1){1}}
\put(1,0.0){\vector(1,0){1}}
\put(2,0.0){\vector(0,1){1}}
\put(2,1.0){\vector(0,1){1}}
\put(0.5,-0.1){\scriptsize \( x \) }
\put(1.5,2.0){\scriptsize \( y \) }
\put(-0.4,-2.1){\scriptsize \( y \) }
\end{picture}}}
\newcommand{\blinkbb}{\mbox{
\begin{picture}(2.5,2.5)
\thicklines
\multiput(0,0.0)(0,-1){3}{\circle*{0.1}}
\multiput(1,1.0)(0,1){2}{\circle*{0.1}}
\multiput(1,0.0)(0,){1}{\circle{0.2}}
\multiput(2,2)(0,1){1}{\circle*{0.1}}
\put(0,-1.0){\vector(0,-1){1}}
\put(0,0){\vector(0,-1){1}}
\put(1,0){\vector(-1,0){1}}
\put(1,0){\vector(0,1){1}}
\put(1,1){\vector(0,1){1}}
\put(1,2){\vector(1,0){1}}
\put(1.25,-0.1){\scriptsize \( x \) }
\put(2.2,2.0){\scriptsize \( y \) }
\put(-0.4,-2.1){\scriptsize \( y \) }
\end{picture}}}
\newcommand{\blinkbc}{\mbox{
\begin{picture}(2.5,2.5)
\thicklines
\multiput(1,1)(0,1){2}{\circle*{0.1}}
\multiput(1,0)(0,0){1}{\circle{0.2}}
\multiput(2,0)(0,-1){3}{\circle*{0.1}}
\multiput(0,2)(0,1){1}{\circle*{0.1}}
\put(1,2){\vector(-1,0){1}}
\put(1,1){\vector(0,1){1}}
\put(1,0){\vector(0,1){1}}
\put(1,0){\vector(1,0){1}}
\put(2,0){\vector(0,-1){1}}
\put(2,-1){\vector(0,-1){1}}
\put(0.5,-0.1){\scriptsize \( x \) }
\put(1.5,-2.1){\scriptsize \( y \) }
\put(-0.5,2.0){\scriptsize \( y \) }
\end{picture}}}
\newcommand{\blinkbd}{\mbox{
\begin{picture}(2.5,2.5)
\thicklines
\multiput(0,0)(0,1){3}{\circle*{0.1}}
\multiput(1,-1)(0,-1){2}{\circle*{0.1}}
\multiput(1,0)(0,-1){1}{\circle{0.2}}
\multiput(2,-2)(0,1){1}{\circle*{0.1}}
\put(0,1){\vector(0,1){1}}
\put(0,0){\vector(0,1){1}}
\put(1,0){\vector(-1,0){1}}
\put(1,0){\vector(0,-1){1}}
\put(1,-1){\vector(0,-1){1}}
\put(1,-2){\vector(1,0){1}}
\put(1.25,-0.1){\scriptsize \( x \) }
\put(2.1,-2.1){\scriptsize \( y \) }
\put(-0.5,2.0){\scriptsize \( y \) }
\end{picture}}}
\newcommand{\staple}{\mbox{
\begin{picture}(1.2, 2.2)
\thicklines
\put(0,1.2){\vector(0,1){1}}
\put(0,1){\vector(0,-1){1}}
\put(0,2.2){\vector(1,0){1}}
\put(0,0.){\vector(1,0){1}}
\put(1,0.){\vector(0,1){1}}
\put(1,2.2){\vector(0,-1){1}}
\end{picture}}}
\newcommand{\staplea}{\mbox{
\begin{picture}(1.2, 2.2)
\thicklines
\put(1,1.2){\vector(0,1){1}}
\put(0,2.2){\vector(0,-1){1}}
\put(1,2.2){\vector(-1,0){1}}
\put(1,1){\vector(0,-1){1}}
\put(1,0){\vector(-1,0){1}}
\put(0,0){\vector(0,1){1}}
\end{picture}}}
\newcommand{\link}{\mbox{
\begin{picture}(1.1, .1)
\thicklines
\put(0,0.1){\vector(1,0){1}}
\end{picture}}}
\newcommand{\linka}{\mbox{
\begin{picture}(1.1, .1)
\thicklines
\put(1,0.1){\vector(-1,0){1}}
\end{picture}}}

\begin{document}
\thispagestyle{empty}
%
 \mbox{} \hfill BI-TP 2000/41\\
\begin{center}
{{\large \bf Quark Mass and Flavour Dependence of the QCD\\[3mm]
Phase Transition}
 } \\
\vspace*{1.0cm}
{\large F. Karsch, E. Laermann and A. Peikert}

\vspace*{1.0cm}
{\normalsize
$\mbox{}$ {Fakult\"at f\"ur Physik, Universit\"at Bielefeld,
D-33615 Bielefeld, Germany}
}
\end{center}
\vspace*{1.0cm}
\centerline{\large ABSTRACT}

\baselineskip 20pt

\noindent
We analyze the quark mass and flavour dependence of the QCD phase 
transition temperature. When the lightest pseudo-scalar meson mass
($m_{\rm PS}$) is larger than 2~GeV the critical temperature is 
controlled by the gluonic sector of QCD alone. For smaller values of 
the lightest meson mass the pseudo-critical temperature decreases 
slowly with $m_{\rm PS}$. For a large regime of meson masses
the pseudo-critical temperature of 2-flavour QCD is about 10\%
larger than in the 3-flavour case. On lattices with temporal extent
$N_\tau=4$ an extrapolation to the chiral limit yields 
$T_c = (173\pm 8)$~MeV and $(154 \pm 8)$~MeV for 2 and 3-flavour QCD, 
respectively. 
We also analyze dynamical quark mass effects on the screening of the
heavy quark potential. A detailed analysis of the heavy quark
free energy in 3-flavour QCD shows that close to $T_c$ screening
effects are approximately quark mass independent already for 
pseudo-scalar meson masses $m_{\rm PS}\; \lsim \; 800$~MeV
and screening sets in at distances $r \simeq 0.3$~fm. 
\vfill
\noindent
 \mbox{}December 2000\\
\eject
\baselineskip 15pt

\section{Introduction}

One of the most prominent non-perturbative features of QCD is the existence 
of a chiral symmetry restoring and deconfining phase transition at high 
temperature. Details of this phase transition, the quantitative value
of the transition temperature as well as the order of the transition,
will depend on basic parameters of QCD, {\it i.e.} the number of 
flavours ($n_F$) and the values of the quark masses ($m_q$). The influence 
of global symmetries of the QCD Lagrangian 
on the order of the phase transition is well understood 
and predictions in the chiral \cite{Pis84}
as well as quenched \cite{Yaffe} limit have to a large extent been verified 
by lattice calculations. It has been shown that the chiral transition
is first order in 3-flavour QCD \cite{old3} and likely to be continuous in 
the 2-flavour case \cite{Lae98}. The explicit flavour as well as 
quark mass dependence of the transition temperature itself, however,
is not related to universal properties and therefore is less well understood. 

Of great interest is, of course, a quantitative determination of
the QCD transition temperature\footnote{In the following we often will 
talk about the {\it QCD transition} without distinguishing between a 
rapid crossover and a true phase transition. This should be clear from 
the context.} in the limit of light quark masses.
The transition temperatures in 2 and 3-flavour QCD will provide
upper and lower limits for the transition temperature of QCD
with two light and a heavier strange quark mass. 
Previous calculations of the transition temperature in QCD with 
dynamical quarks have shown that it is significantly smaller than
the critical temperature determined for the deconfinement transition
which occurs in the limit of infinitely heavy quarks, {\it i.e.} in the 
SU(3) gauge theory \cite{Kar99}. The quark mass dependence of the 
QCD transition temperature, however, is not well understood.
In particular, we would like to understand in more detail what sets 
the scale for the QCD transition temperature. 
It is obvious that for
large quark masses the valence quark dominated hadronic sector will
completely decouple and thermodynamics will be controlled by the gluonic 
sector alone. In the chiral limit, on the other hand, the light pseudo-scalar 
mesons are expected to play a dominant role. 
For instance, studies within the context of $\sigma -$models suggest a
strong dependence of the transition temperature on the pseudo-scalar 
meson mass \cite{sigma}. For realistic quark mass values
the transition temperature itself will be compatible with the value 
of the pseudo-scalar mass and one may expect that the  
rich hadronic resonance spectrum will play an important role in this 
regime \cite{hagedorn}. The importance of a rapidly (exponentially)
rising resonance mass spectrum for the location of the transition temperature
does become apparent already in the $SU(N)$ gauge theories where the
transition temperature is rather low ($T_c \simeq 270$~MeV) despite  
the fact that the lightest excitations in the low temperature phase   
are glueballs with a mass of about 1.5~GeV \cite{glueballs}. In fact, 
the value of the phase transition temperature of $SU(N)$ gauge theories
can be well understood in terms of the exponentially rising mass spectrum 
obtained in string models \cite{string}.
In order to better understand QCD thermodynamics and in particular the
critical behaviour close to the transition temperature a detailed analysis 
of the quark mass dependence of the QCD transition is needed. This 
problem will be addressed here. 
In order to be able to discuss the quark mass dependence of the QCD transition
temperature we also have to address the question in how far
we can compare transition temperatures when the quark masses and/or
the number of flavours are varied. We need an appropriate observable to 
set the scale for the transition temperature which itself is little
influenced by these external parameters.
In order to reach quantitative conclusions
on these questions it will be mandatory to get the intrinsic systematic
lattice effects, cut-off dependence and finite volume effects, under
control. We therefore will use here improved gauge and fermion actions
for our analysis of the quark mass and flavour dependence of the
transition temperature in QCD.

We also discuss the influence of dynamical quarks on the heavy quark 
potential, or more precisely the heavy quark free energy extracted at
finite temperature from Polyakov loop correlation functions. In the 
case of 3-flavour QCD we analyze in detail the quark mass and 
temperature dependence of the screening of the heavy quark free energy,
which becomes significant already below $T_c$.

This paper is organized as follows. In the next section we will
introduce the improved gauge and staggered fermion actions used
for our simulations and give some details on the simulation parameters.
In Section 3 we present our results on the flavour and quark mass 
dependence of the transition temperature and give an estimate for
the critical temperatures of the chiral transition in massless
2 and 3-flavour QCD. In Section 4 we discuss in more detail the
quark mass and temperature dependence of the heavy quark free
energy. We finally give our conclusions in Section 5.

\section{Thermodynamics with an improved staggered fermion action}

The use of improved gauge and fermion actions is mandatory in
thermodynamic calculations which generally are performed at quite
large values of the lattice spacing, $a$, because one has to use quite
small values for the temporal extent of the lattice. The finite cut-off 
effects become particularly striking in the calculation of bulk thermodynamic
observables. In fact, it has been shown just in these cases that improved 
actions can be particularly efficient and 
reduce the cut-off dependence of thermodynamic observables significantly
\cite{Kar99}. 

In the pure gauge sector the tree level, Symanzik improved action 
yields a satisfactory description of the high temperature ideal gas limit 
of QCD already on lattices with temporal extent $N_\tau =4$
\cite{Bei99}. Similarly good results can be achieved in the fermion
sector with improved staggered fermions where three-link terms are added
to the ordinary one link discretization of the first order derivatives
appearing in the fermion Lagrangian \cite{Hel99,Kar00}. 

In the present analysis we will use such an improved action,
which in addition to the standard Wilson plaquette term and the standard
1-link staggered fermion action also includes the planar 6-link Wilson 
loops and bended 3-link terms which improve the rotational symmetry of
the staggered fermion action \cite{Kar00}. The partition function  
for QCD with $n_f$ quark flavours reads 
\begin{equation}
Z(T,V) = \int \prod_{x,\mu}{\rm d} U_{\mu}(x) {\rm e}^{-\beta S_G }
\prod_{q=1}^{n_f}\biggl(\int \prod_x {\rm d}\bar{\chi}_x
{\rm d}\chi_x\; {\rm e}^{-S_F(m_{q})} \biggr)^{1/4}
\label{partition} 
\end{equation}
where $x$ labels the sites on a hypercubic lattice of size
$N_\sigma^3\times N_\tau$. 
The gauge ($S_G$) and fermion ($S_F$) actions are given
by
\begin{eqnarray}
S_G &=& c_4\; S_{plaquette} + c_6\; S_{planar} \nonumber \\
&\equiv& \sum_{x, \nu > \mu}~ {5\over 3}~\left(
1-\frac{1}{3}\re\tr\plaq_{\mu\nu}(x)\right) \nn\\
& & -{1\over 6 }\left(1-\frac{1}{6}\re\tr
\left(\loOp_{\mu\nu}(x)+\lOop_{\mu\nu}(x)\right)\right)
\label{sg}
\end{eqnarray}
\begin{eqnarray}
\lefteqn{ S_F (m_{q}) ~=~ c_1^F S_{1-link,fat} (\omega) +
c_3^F S_{3-link}+  m_{q}\sum_x
\bar{\chi}_x^f \chi_x^f \nn ~~~~~~~~~~~~~~~~~~~~~~~}\\
&\equiv& \sum_x \bar{\chi}_x^f~\sum_\mu ~ \eta_\mu(x) ~ \Bigg(
{3\over 8}~\Bigg[ \alink~ +~ \omega~~\sum_{\nu \ne \mu}~~ \alinkfat\Bigg]
\nn \\[4mm]
& & + {1\over96}~\sum_{\nu\ne \mu} ~\Bigg[ \blinkbd + \blinkbc ~+
 \blinkba + \blinkbb \Bigg] \Bigg) \chi_y^f \nn \\[13mm]
& & + m_{q}  \sum_x~\bar{\chi}_x^f \chi_x^f  \quad .
\label{sf}
\end{eqnarray}
Here $\eta_\mu(x) \equiv (-1)^{x_0+..+x_{\mu-1}}$ denotes the
staggered fermion phase factors. Furthermore, we have made explicit 
the dependence of the fermion action on different
quark flavours $q$, and the corresponding bare quark masses $m_q$,
and give an intuitive graphical representation of the action.
The tree level coefficients $c_1^F$ and $c_3^F$ appearing in $S_F$
have been fixed by demanding rotational invariance
of the free quark propagator at ${\cal O}(p^4)$  (``p4-action'')
\cite{Hel99}. 
In addition the 1-link term of the fermion action has been modified by 
introducing ``fat'' links \cite{Blu97} with a weight $\omega = 0.2$.
The use of fat links does lead to a reduction of the flavour symmetry
breaking close to $T_c$ and at the same time does not modify the
good features of the p4-action at high temperature, {\it i.e.} it has 
little influence on the cut-off dependence of bulk thermodynamic 
observables in the high temperature phase \cite{Hel99,Kar00}. 
Further details on the definition of the action are given 
in \cite{Hel99}.

For simplicity, in the following we will refer to the p4-action with a 
fat 1-link term combined with the tree level improved gauge action as 
{\it the p4-action}.

Numerical simulations with dynamical fermions have been performed using
the Hybrid R algorithm \cite{Got87}. This algorithm introduces
systematic errors which depend quadratically on the step size $\delta \tau$ 
used for the integration 
of molecular dynamics trajectories. We have verified on a small
$8^3\times 4$ lattice that the systematic errors with our improved
action are similar in magnitude to those found for the standard action
\cite{Aoki98}. In our simulations we use a step size 
$\delta\tau = \min\{0.4\; m_q, 0.1 \}$. The length of the
molecular dynamics trajectories has been chosen to be $\tau = 0.8$.

\section{Critical Temperature in 2, 2+1 and 3 Flavour QCD}

In this section we want to discuss our results for the pseudo-critical
temperatures obtained for 2 and 3-flavour QCD with different quark masses.
In addition to the results obtained with identical quark mass values
for all flavours we also will present the result of a calculation for
(2+1)-flavour QCD, {\it i.e.} for QCD with two light quark flavours 
($m_{u,d}=0.1$) and one heavier flavour ($m_s =0.25$). Of course,
these bare quark mass parameters receive a multiplicative renormalization. 
For an easier comparison with the better known parameters of the unimproved 
staggered fermion action we note that the bare quark parameter, $m_q$, 
of the p4-action corresponds to about twice the bare quark mass in the
unimproved staggered action. 

The calculations with degenerate quark masses cover a range of pseudo-scalar 
meson mass from about 250~MeV up to the quenched limit of infinitely heavy 
mesons. 
We will analyze the quark mass dependence of the pseudo-critical
temperature and discuss its extrapolation to the chiral limit.
This involves two steps: First of all one has to determine the
pseudo-critical couplings, $\beta_c(m_q)$, on finite temperature
lattices of size $N_\sigma^3 \times N_\tau$. One then has to calculate
physical observables at zero temperature which allow to extract the 
lattice spacing at $\beta_c(m_q)$ through which the temperature is 
fixed, $T_c=1/N_\tau a(\beta_c)$. 

\subsection{Critical couplings on $16^3\times 4$ lattices}

\begin{table}[!t]
\renewcommand{\arraystretch}{1.3}
\center{
\begin{tabular}{|c||r|r|r|l|}
\hline
\multicolumn{5}{|c|}{$n_f=2$}\\
\hline
$m_{u,d}$ & $N_\sigma^3\times N_\tau$ & \#$\beta$ & \#{iter.}&
\multicolumn{1}{|c|}{$\beta_c$} \\
\hline\hline
0.025 & $8^3\times4$ & 3  & 3200   & 3.542~(7) \\
\hline
0.05 & $8^3\times4$  & 4  & 3700   & 3.585~(11) \\
\hline
0.10 & $16^3\times4$ & 10 & 36300  & 3.646~(4)  \\
\hline
0.20 & $8^3\times4$  & 6  & 6300   & 3.778~(12) \\
\hline\hline
\multicolumn{5}{|c|}{$n_f=2+1$}\\
\hline
$m_{u,d}/m_{s}$ & $N_\sigma^3\times N_\tau$ & \#$\beta$ & \#{iter.}&
\multicolumn{1}{|c|}{$\beta_c$}  \\
\hline\hline
0.10/0.25 & $16^3\times4$ & 10 & 31550 & 3.543~(2) \\
\hline\hline
\end{tabular}
\caption{
\label{tab:crit_couplings}
Simulation parameters for the determination of the critical temperature 
for 2 and $2+1$ flavours. The largest lattice size simulated, the
number of $\beta$ values and iterations as well as the finally determined 
pseudo-critical couplings are given. 
}
}
\end{table}

The finite temperature calculations have been performed on lattices
with temporal extent $N_\tau=4$ and spatial lattice sizes $N_\sigma =
8$, 12 and 16. For each value of the quark mass we have performed 
simulations at several values of the gauge coupling $\beta$.
Details on the simulation parameters and the statistics
collected on our largest lattices are given in
Table~\ref{tab:crit_couplings}.
For the determination of the pseudo-critical couplings we calculated the
average Polyakov loop,
\begin{equation}
L \equiv {1 \over N_\sigma^3} \left| \sum_{\vec{x}} L(\vec{x})\right| 
={1 \over N_\sigma^3} \left| {1\over 3} \sum_{\vec{x}} 
{\Tr}\prod_{x_4=1}^{N_\tau} U_4(\vec{x},x_4) \right| \quad ,
\label{polyakov}
\end{equation}
and used a noisy estimator for the chiral condensate which is based
on 25 random vectors $R_i$, 
\begin{equation}
\overline{\psi}\psi =
\frac{1}{N_\tau N_\sigma^3}\frac{n_f}{4} \Tr \; M^{-1} \simeq
{n_f\over 100 N_\tau N_\sigma^3} \sum_{i=1}^{25} R_i^{\dagger}M^{-1} R_i
\quad .
\label{condensate}
\end{equation}
As usual we have introduced here the fermion matrix $M$, which is 
defined through the matrix form of Eq.~\ref{sf}, {\it i.e.}
$S_F \equiv \sum_{q=1}^{n_f} \bar{\chi}\; M(m_q) \; \chi$.
At each value of the quark mass
the data from runs at all $\beta$-values have been used in a 
Ferrenberg-Swendsen reweighting \cite{Fer88} to determine the maxima in
the related Polyakov-loop and chiral susceptibilities\footnote{Note
that we only calculate the disconnected part of the complete chiral 
susceptibility,
$\chi_{\overline{\psi}\psi} \sim \partial^2 \ln Z/ \partial (m_q)^2 $.},
\begin{eqnarray}
\chi_L &\equiv&  N_\sigma^3 \left(\left<  L^2\right> - 
\left< L \right>^2 \right)~~~, \nonumber \\
\chi_{\overline{\psi}\psi, {\rm disc}}&=&
\frac{1}{N_\tau N_\sigma^3}\frac{n_f}{16}
\left( \left< \left( \Tr \; M^{-1}\right)^2 \right>- 
\left< \Tr \; M^{-1}\right>^2 \right)  
\label{susceptibilities}
\end{eqnarray}
as well as the susceptibility of the gauge action,
\begin{equation}
\chi_S \equiv  N_\sigma^3N_\tau \left(\left<  S_G^2\right> -
\left< S_G \right>^2 \right)~~.
\end{equation}
At intermediate values of the quark mass the transition
to the high temperature phase of QCD is not related to any form of 
critical behaviour in thermodynamic observables. Nonetheless, the 
 maxima of the susceptibilities define pseudo-critical couplings, 
$\beta_c(m_q)$, for the transition to the high temperature phase of QCD, which 
manifests itself in 
a sudden change of thermodynamic observables in a small temperature
interval\footnote{It recently has
been suggested that the sudden change in behaviour at a certain
pseudo-critical temperature may be characterized in terms of a non-thermal,
percolation phase transition \cite{satz}.}. For small (vanishing) and large 
values of the quark mass the transition, however, becomes a true phase 
transition in the infinite volume limit. In these cases the pseudo-critical 
couplings 
give finite volume estimates for the location of the critical point in the 
thermodynamic limit.  

\begin{table}[!th]
\vspace{-2.0truecm}
\renewcommand{\arraystretch}{1.3}
\center{
\begin{tabular}{|c|r|r|r||l|l|l|}
\hline
\multicolumn{4}{|l||}{}&\multicolumn{3}{c|}{pseudo-critical couplings $\beta_c$}\\
\hline
$m_q$ & $N_\sigma$ & \# $\beta$  &  \# iter.  &\multicolumn{1}{c|}{\
  max($\chi_S$)} &\multicolumn{1}{c|}{ max($\chi_L$)} &\multicolumn{1}{c|}{\
  max($\chi_{\bar{\psi}\psi,disc}$)}\\
\hline\hline
\multirow{2}{5mm}{0.01}
 & 8  & 9 & 10300  & 3.2865~(98) & 3.2887~(65) & 3.2841~(57) \\\cline{2-7}
 & 16 & 3 & 4440  & 3.2784~(89) & 3.2804~(37) & 3.2806~(20) \\\cline{2-7}
\hline\hline
\multirow{2}{9mm}{0.025}
 & 8  & 8  & 14000 & 3.3265~(93) & 3.3318~(24)  & 3.3281~(25) \\\cline{2-7}
 & 16 & 6  &  8200 & \multicolumn{1}{c|}{-}  & 3.3428~(159)  & 3.3194~(21) \\
\hline\hline
\multirow{2}{5mm}{0.05}
 & 8  & 11 & 16250 & \multicolumn{1}{c|}{-}  & 3.4018~(35)  & 3.3930~(201) \\\cline{2-7}
 & 16 &  4 &  4500 & 3.3866~(47) & 3.3950~(150)  & 3.3877~(26)  \\
\hline\hline
\multirow{2}{5mm}{0.10}
 & 8  & 12 & 25100 & 3.4637~(34) & 3.4864~(25)  & 3.4842~(151) \\\cline{2-7}
 & 16 &  9 & 15350 & 3.4752~(10) & 3.4856~(138) & 3.4756~(14)  \\\cline{2-7}
\hline \hline
\multirow{3}{5mm}{0.20}
 & 8  & 16 & 39250  & 3.5184~(462) & 3.6216~(27) & 3.6204~(32) \\\cline{2-7}
 & 12 &  5 & 18700  & 3.5986~(25)  & 3.6020~(32) & 3.6004~(18) \\\cline{2-7}
 & 16 &  8 & 29650  & 3.5998~(14  )& 3.6017~(17) & 3.6000~(33) \\
\hline\hline
\multirow{3}{5mm}{0.40}
 & 8  & 11 & 36600  & 3.7756~(64)  & 3.7848~(23) & 3.7808~(27) \\\cline{2-7}
 & 12 &  3 & 11700  & 3.7810~(24)  & 3.7781~(17) & 3.7797~(16) \\\cline{2-7}
 & 16 &  5 & 16250  & 3.7653~(16)  & 3.7719~(33) & 3.7746~(40) \\
\hline\hline
\multirow{3}{5mm}{0.60}
 & 8  & 12 & 33100  & 3.7983~(159) & 3.8865~(17) & 3.8806~(12) \\\cline{2-7}
 & 12 &  4 & 16500  & 3.8718~(47)  & 3.8770~(16) & 3.8798~(38) \\\cline{2-7}
 & 16 &  6 & 41750  & 3.8210~(233) & 3.8766~(6)  & 3.8764~(6)  \\
\hline\hline
\multirow{1}{5mm}{0.70}
 & 16  &  4 & 24000  & 3.7983~(159) & 3.8865~(17) & 3.8806~(12) \\\cline{2-7}
\hline\hline
\multirow{1}{5mm}{0.80}
 & 16  &  3 & 25300  & 3.9358~(18)  & 3.9377~(10) & 3.9368~(14) \\\cline{2-7}
\hline\hline
\multirow{3}{5mm}{1.00}
 & 8  &  8 &  21750 & 3.9680~(6)   & 3.9686~(27) & 3.9663~(11) \\\cline{2-7}
 & 12 &  5 &  24520 & 3.9717~(39)  & 3.9712~(22) & 3.9711~(15) \\\cline{2-7}
 & 16 &  8 & 101300 & 3.9771~(15)  & 3.9778~(10) & 3.9808~(16) \\
\hline\hline
\multirow{5}{5mm}{$\infty$}
 & 8  & 10 & 129000 & 4.0554~(27)  & 4.0745~(17) & \multicolumn{1}{c|}{-} \\\cline{2-7}
 & 12 & 14 & 154900 & 4.0708~(27)  & 4.0708~(6)  & \multicolumn{1}{c|}{-} \\\cline{2-7}
 & 16 & 8  & 157500 & 4.0719~(7)   & 4.0715~(5)  & \multicolumn{1}{c|}{-} \\\cline{2-7}
 & 24 & 3  &  75200 & 4.0724~(4)   & 4.0722~(4)  & \multicolumn{1}{c|}{-} \\\cline{2-7}
 & 32 & 1  &  46800 & 4.0733~(4)   & 4.0729~(3)  & \multicolumn{1}{c|}{-} \\\cline{2-7}
 & $\infty$ & - & - & 4.0732~(4)   & 4.0730~(3)  & \multicolumn{1}{c|}{-} \\
\hline
\end{tabular}
\caption{
\label{tab:betac_nf3}
Pseudo-critical couplings of the p4-action calculated on lattices
with spatial extent $N_\sigma =8$, 12 and 16 and temporal extent $N_\tau=4$ 
for 3-flavour QCD with different quark masses. 
Errors on the maxima of the susceptibilities have been
determined from a jackknife analysis.  
}
}
\end{table}

Previous analyses of the volume dependence of the pseudo-critical couplings
performed with unimproved gauge and staggered fermion actions 
\cite{Lae98,Aoki98}
have shown that finite size effects are small at intermediate values
of the quark mass. Moreover, the pseudo-critical couplings seem to show little 
dependence on the observable used to define them. This is also supported by
our analysis of finite volume effects in simulations with the  p4-action. 
In the case of 3-flavour QCD we have determined the pseudo-critical 
couplings $\beta_c(m_q)$ for 10 different values of the bare quark 
mass in the interval $0.01\le m_q \le 1.0$ on lattices of spatial extent 
$N_\sigma = 8$,~12 and 16. The results of this analysis are summarized
in Table~\ref{tab:betac_nf3}. Within the statistical errors of the present
analysis we do not observe any clear volume dependence at intermediate 
values of the quark mass and even not for the smallest value, $m_q=0.01$, 
considered here. This, however, is no longer true for large 
masses, where the onset of the first order deconfining transition leads to
large correlation lengths. This is reflected in a strong volume dependence of 
the peak height of the susceptibilities and a visible volume dependence
of the pseudo-critical couplings. 
Note that the location of the peak of the action susceptibility $\chi_S$ 
shows the strongest volume
dependence. However, on the largest lattice ($16^3\times 4$) it also agrees 
within statistical errors with all the other observables used to locate
the pseudo-critical points. In particular, we do not find any systematic
volume dependence in the peak location of 
$\chi_L$ and $\chi_{\bar{\psi}\psi,{\rm disc}}$ for
quark masses $m_q\le 0.6$.
The pseudo-critical couplings taken from the largest lattice which have
been used for the determination of the critical temperature are 
also given in Table~\ref{tab:crit_couplings}.

The extrapolation of the pseudo-critical couplings to the chiral limit
will depend on the order of the phase transition which, in accordance 
with present knowledge from lattice calculations, 
is expected to be second order for $n_f=2$ and first order for $n_f\ge 3$
\cite{Pis84}.
If the transition is first order in the chiral limit it will remain
to be first order also for small values of the quark mass. The shift 
in the critical coupling will then depend linearly on the quark
mass with a slope controlled by the zero quark mass discontinuities in 
the order parameter and gauge action. From a Taylor expansion
of the free energy below and above the critical point one 
obtains\footnote{This follows 
the same line of arguments given in Ref.~\cite{Has83} for the quark mass
dependence of the deconfinement transition in the limit of heavy quarks.},
\begin{equation}
\beta_c(m_q) = \beta_c(0) + 
{\Delta \langle \bar{\psi}\psi \rangle \over \Delta \langle S_G \rangle}
m_q 
\qquad , \qquad n_f \ge 3~~.
\end{equation}
In the case of a second order transition, however, the diverging correlation
length at the critical point will lead to a stronger variation of
the critical couplings with the quark mass. From the non-analytic structure
of the singular part of the free energy one finds that the critical couplings 
will scale like,
\begin{equation}
\beta_c(m_q) = \beta_c(0) + c\cdot (m_q)^{1/\beta\delta} \qquad , \qquad
n_f=2~~.
\label{fitnf2}
\end{equation}
Here $1/\beta\delta$ is a combination of critical exponents which for the 
case of 2-flavour QCD are expected to be those of the three
dimensional $O(4)$ symmetric spin models, {\it i.e.} in the continuum limit
one expects to find $1/\beta\delta \simeq 0.55$. Of course, the regular
part of the free energy will also add a linear dependence on $m_q$ to
this leading order behaviour.
\begin{figure}
\begin{center}
\epsfig{file= 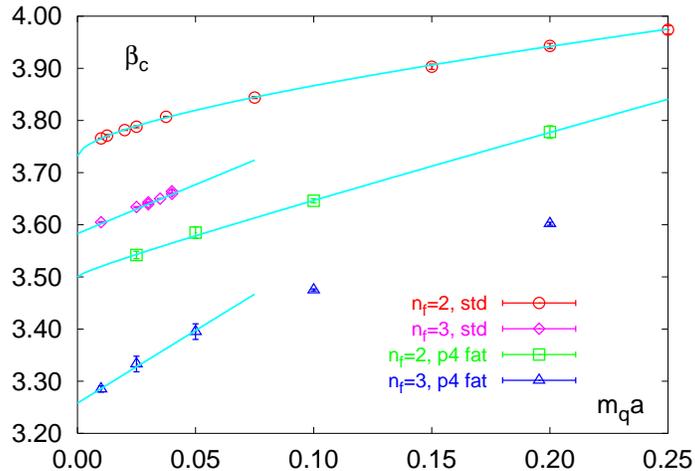,width=95mm}
\end{center}
\caption{
\label{fig:betac}
Pseudo-critical couplings of 2 and 3 flavour QCD versus bare quark 
mass calculated on lattices of size $16^3\times 4$ with the standard Wilson
plaquette and standard staggered fermion action (std) and the 
improved gauge and fermion action (p4) described in section 2.
The data for the standard action have been shifted by $\Delta\beta= -1.5$.
Lines show linear fits to the 3-flavour data for $m_q \le 0.05$ and
``O(4)+linear'' fits for the 2-flavour data with $m_q \le 0.2$.} 
\end{figure}

In Figure~\ref{fig:betac} we show our results for $\beta_c(m_q)$ in 2 and 
3-flavour QCD obtained for small and intermediate values of the quark mass
with the p4-action. Furthermore, we show results 
from calculations with the standard Wilson gauge and standard staggered 
fermion action \cite{Aoki98,Aoki99,Luetgemeier1998,Got87b,Fuk90} 

The similarity in the quark mass dependence of $\beta_c(m_q)$ seen 
in Figure~\ref{fig:betac} for the 2 and 3-flavour cases suggests that 
for $n_f$=2 also the subleading corrections, linear in $m_q$ are still 
important in the quark mass regime explored by present simulations.
We thus have extrapolated the critical couplings
to the chiral limit by assuming the $O(4)$ symmetric ansatz given
in Eq.~\ref{fitnf2} and allowing for a subleading term linear in $m_q$.
This yields for

\vspace{0.2cm}
\noindent
\underline{$n_f=2$:}
\begin{equation}
\beta_c(0) = \cases{ 5.232\pm 0.002 & standard gauge and fermion action
\cr
3.48 \pm 0.03 & improved gauge and fermion action }
\end{equation}
The result for the standard staggered action 
is consistent with fits solely based on the ansatz given
in Eq.~\ref{fitnf2} with free exponent $1/ \beta\delta$ \cite{Aoki98,Kar94}
as well as fits in which the exponent has been fixed to its $O(4)$ value, 
{\it i.e.} $1/ \beta\delta = 0.55$.

In the 3-flavour case, on the other hand, the linear dependence on $m_q$ is
evident for small quark masses and linear fits with a small $\chi^2/(d.o.f.)$
can be performed for $m_q\le 0.05$. From these fits we find for
the critical couplings of the 3-flavour theory in the chiral limit,

\vspace{0.2cm}
\noindent
\underline{$n_f=3:$}
\begin{equation}
\beta_c(0) = \cases{ 5.083\pm 0.004 & standard gauge and fermion action \cr
3.258 \pm 0.004 & improved gauge and fermion action }
\end{equation}

The quark mass dependence of $\beta_c(m_q)$ is significantly different
in the case of improved and unimproved actions.
From the fits of the critical couplings we find,
\begin{equation}
{\Delta \langle \bar{\psi}\psi\rangle \over \Delta\langle S_G \rangle 
\cdot N_\tau} = 
\cases{ 0.49\pm 0.01 & standard action \cr
0.70 \pm 0.04 & improved action }
\label{slope}
\end{equation}
However, in order to relate these slope parameters 
to a physical observable, the ratio of the discontinuity in the chiral
condensate and the latent heat $\Delta \epsilon$, on still has
to determine a normalization constant $Z_m(\beta_c)$ \cite{Gon82} that
turns $\langle \bar{\psi}\psi\rangle$ into a renormalization group invariant 
observable,
\begin{equation}
{T_c\cdot \Delta (\bar{\psi}\psi)^{RGI} \over \Delta \epsilon}  =
{\Delta \langle \bar{\psi}\psi\rangle \over \Delta\langle S_G \rangle
\cdot N_\tau}Z_m^{-1}(\beta_c)\quad . 
\end{equation}
This may partly account for the observed difference between the
standard and improved actions.
 
\subsection{Setting the scale at zero temperature} 

In order to relate the pseudo-critical couplings to physical 
temperatures and to perform the extrapolation to the chiral limit 
we have performed zero temperature simulations on lattices of size 
$16^4$. In addition to pseudo-scalar ($m_{\rm PS}$) and vector ($m_V$) 
meson masses we also have calculated the string tension ($\sigma$) 
from large Wilson loops. 

The spectrum calculations performed at $\beta_c(m_q)$ follow the 
conventional approach. They are based on 
wall source operators as described e.g. in Ref.~\cite{Alt93}. Masses have
then been extracted from the exponential decay of correlation functions
constructed from wall source and local sink operators. Also the 
calculation of the string tension is based on a standard approach. We
use smeared Wilson loops \cite{ape87} from which we extract a potential
which then is fitted to an ansatz which includes a Coulomb and a
linearly rising term. The notion of a string tension in the presence of 
dynamical quarks does, however, require a bit more explanation.
In the presence of dynamical quarks the heavy quark potential is no 
longer linear at large distances as the spontaneous creation of quark 
anti-quark pairs from the vacuum will lead to string breaking. 
Nonetheless, it has been found that the ``potential'' extracted from Wilson 
loops does not show evidence for string breaking\footnote{
String breaking has been found to occur at non-zero temperature where 
the heavy quark free energy 
can be extracted from Polyakov loop correlation functions \cite{DeT99}.
We will discuss this in more detail in Section 4.} 
even at distances of up to about $2{\rm fm}$ \cite{wilsonstring}. 
Wilson loops thus can still be used 
to extract a ``string tension'' from its large distance 
behaviour. As will become clear in the following a 
comparison with quenched and partially quenched spectrum calculations shows
that this string tension shows only little quark mass and flavour 
dependence.

The meson masses 
and the string tension calculated at $\beta_c (m_q)$ for 2 and
3-flavour QCD are given in Table~\ref{tab:crit_temp}. Here we also
give the result from a simulation of (2+1)-flavour QCD.
Furthermore, for $n_f = 3$ we have performed calculations with bare quark
mass $m_q =0.1$ at various $\beta$-values that cover the region
of critical couplings obtained with small quark masses. We will
use these data in the next section to determine the vector meson mass
($m_\rho$) in the chiral limit at $\beta_c (m_q=0)$.

\begin{table}[t]
\renewcommand{\arraystretch}{1.3}
\center{
\begin{tabular}{|c||l|l|l|l|l|l|}
\hline
\multicolumn{7}{|c|}{$N_f=2$}\\
\hline
$m_{u,d}$ & \multicolumn{1}{|c|}{$\beta_c$} & \multicolumn{1}{|c|}{$\sigma a^2$} & 
\multicolumn{1}{|c|}{$m_{\rm PS} a$} & \multicolumn{1}{|c|}{$m_{\rm V} a$} & 
\multicolumn{1}{|c|}{$T_c/\sqrt{\sigma}$} & \multicolumn{1}{|c|}{$T_c/m_{\rm V}$} \\
\hline\hline
0.025 & 3.542~(7)& 0.300~(15) &  0.507~(2) & 1.175~(35) & 0.456~(11)  & 0.213~(6)\\
\hline
0.05 & 3.585~(11) & 0.286~(14) &  0.697~(6) & 1.260~(35) & 0.467~(11)  & 0.198~(6)\\
\hline
0.10 & 3.646~(4)  & 0.271~(10) & 0.958~(2)  & 1.377~(25) & 0.480~(10) & 0.182~(4)\\
\hline
0.20 & 3.778~(12) & 0.205~(12) & 1.295~(15) & 1.530~(35) & 0.552~(16) & 0.163~(4))\\
\hline\hline
\multicolumn{7}{|c|}{$N_f=2+1$}\\
\hline
$m_{u,d}/m_{s}$ & \multicolumn{1}{|c|}{$\beta_c$} &\multicolumn{1}{|c|}{$\sigma a^2$} & 
\multicolumn{1}{|c|}{$m_{\rm PS} a$} & \multicolumn{1}{|c|}{$m_{\rm V} a$} & 
\multicolumn{1}{|c|}{$T_c/\sqrt{\sigma}$} & \multicolumn{1}{|c|}{$T_c/m_{\rm V}$} \\
\hline\hline
0.10/0.25 & 3.543~(2) & 0.271~(11) & 0.962~(3) & 1.343~(20) & 0.480~(10) & 0.186~(3)\\
\hline\hline
\multicolumn{7}{|c|}{$N_f=3$}\\
\hline
$m_{u,d}$ & \multicolumn{1}{|c|}{$\beta_c$}  & \multicolumn{1}{|c|}{$\sigma a^2$} & 
\multicolumn{1}{|c|}{$m_{\rm PS} a$} & \multicolumn{1}{|c|}{$m_{\rm V} a$} & 
\multicolumn{1}{|c|}{$T_c/\sqrt{\sigma}$} & \multicolumn{1}{|c|}{$T_c/m_{\rm V}$} \\
\hline\hline
0.025 & 3.329~(15) & 0.350~(20) & 0.509~(1) & 1.290~(40) & 0.423~(12) &
0.194~(10)\\
\hline
0.05 & 3.395~(15) & 0.303~(13) & 0.706~(6) & 1.320~(35) & 0.454~(10) &
0.189~(6)\\
\hline
0.10 & 3.475~(2) & 0.283~(11) & 0.967~(1) & 1.415~(15) & 0.470~(9) & 0.177~(2)\\
\hline
0.20 & 3.602~(3) & 0.248~(4)  & 1.322~(2) & 1.608~(9)  & 0.502~(4) & 0.155~(1)\\
\hline
0.40 & 3.772~(4) & 0.189~(4)  & 1.814~(4) & 1.985~(10) & 0.575~(6) & 0.126~(1)\\
\hline
0.60 & 3.877~(2) & 0.176~(3)  & 2.210~(4) & 2.347~(12) & 0.596~(5) & 0.107~(1)\\
\hline
1.00 & 3.978~(2) & 0.154~(2)  & 2.838~(6) & 2.979~(15) & 0.637~(4) & 0.084~(1)\\
\hline
\end{tabular}
\caption{
\label{tab:crit_temp}
The string tension, pseudo-scalar masses and vector meson 
calculated at the
critical couplings which are given in the second column. In the last two 
columns we give the critical temperature in terms of the string tension 
and vector meson mass. }
}
\end{table}

In order to quantify the flavour and quark mass dependence of the
QCD transition temperature it is, of course, important to understand
to what extent the zero temperature observables used to set the scale
are dependent on $m_q$ and $n_f$. Clearly the hadron masses will
depend on the bare quark masses $m_q$. With decreasing quark masses
one also might expect some $n_f$-dependence resulting from sea quark
contributions to the hadron masses. Quenched spectrum calculations,
however, suggest that the latter effect is small compared to the
explicit valence quark mass dependence \cite{quenched}. The string tension,
on the other hand, is expected to show a much weaker dependence on 
the quark masses. This is supported by the fact that 
in the quenched limit, {\it i.e.} in the pure $SU(3)$ gauge theory
the value for the string tension deduced from the ratio 
$(\sqrt{\sigma}/m_\rho)_{\rm quenched}$ 
in the limit of vanishing valence quark mass \cite{Wittig}
is consistent with phenomenological values used as input in heavy quark 
spectrum calculations \cite{potential_spectrum}. 

\begin{figure}
\begin{center}
\epsfig{file=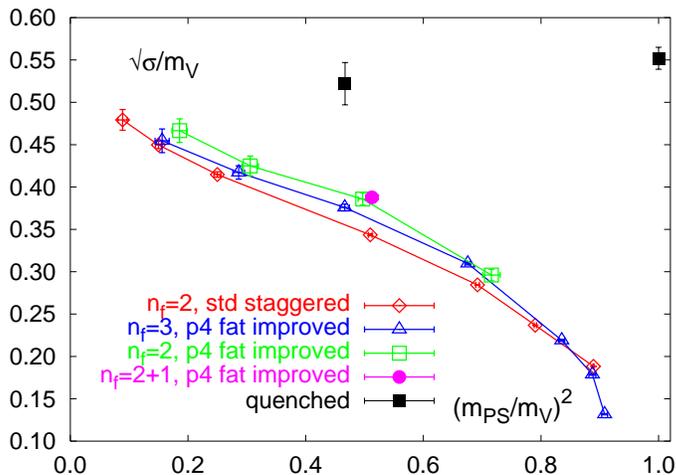,width=95mm}
\end{center}
\caption{
\label{fig:sigma_rho}
Square root of the string tension in units of the 
vector meson mass as a function of the ratio of pseudo-scalar
and vector meson masses. Shown are results for 2, 2+1 and 3 flavour 
QCD obtained with the improved gauge and staggered fermion 
action given in Eqs.~\ref{sg} and \ref{sf} . Also shown are results 
from calculations with unimproved staggered fermions \cite{Luetgemeier1998}.
The black squares show results from quenched calculations and partially
quenched calculations with a sea quark mass of $m_q=0.1$.} 
\end{figure}

In Figure~\ref{fig:sigma_rho} we show the ratio $\sqrt{\sigma}/m_{\rm V}$
as a function of $(m_{\rm PS}/m_{\rm V})^2$. In addition to
the results obtained from simulations with dynamical quarks
at various values of the bare quark mass, we also show in this figure
results obtained in quenched QCD \cite{Wittig} and from a
partially quenched analysis. 
For the latter we 
have performed a spectrum analysis in 3-flavour QCD with valence quarks of mass
$m_q=0.02$ and 0.05 on gauge field configurations generated with dynamical 
quarks of mass $m_q=0.1$ at the critical coupling, $\beta_c(m_q=0.1)=3.475$. 
Results for the partially quenched meson masses obtained in this case
are given in Table~\ref{tab:part_quenched}. The square root of
the string tension in units of the vector meson mass 
in the limit of vanishing valence quark mass is obtained as
\begin{equation}
{\sqrt{\sigma} \over m_\rho} = \cases{0.552 \pm 0.013 &, quenched
\cite{Wittig} \cr
0.532 \pm 0.018 &, partially quenched for $m_q=0.1,~n_f=3$}
\label{phscale}
\end{equation}
This clearly is also consistent with an extrapolation of the 
ratio $\sqrt{\sigma} / m_V$ to the chiral limit which yields 
$\sqrt{\sigma} / m_\rho \simeq 0.50$. Calculated at
non-zero values of the bare quark mass this ratio, however, does 
show a strong dependence on $m_q$.

\begin{table}[t]
\renewcommand{\arraystretch}{1.3}
\center{
\begin{tabular}{|c|c|c|}
\hline
$m_{q}$ & $m_{\rm PS} a$ & $m_{\rm V} a$ \\
\hline\hline
0.02 & 0.448~(2) & 1.08~(1)  \\
\hline
0.05 & 0.695~(2) & 1.23~(2) \\
\hline
\end{tabular}
\caption{
\label{tab:part_quenched}
Partially quenched spectrum results in 3-flavour
QCD performed at $\beta_c(m_q=0.1)$ on $16^4$ lattices. Results for 
$m_q=0.1$ and the corresponding value of the string tension at the 
critical coupling are given in Table~\ref{tab:crit_temp}.}
}
\end{table}

We thus find similar results for the ratio $\sqrt{\sigma}/m_\rho$
in the quenched and in the chiral limit. This suggests
that the string tension extracted from Wilson loops as 
well as partially quenched meson masses depend only weakly on 
the number of flavours and the value of the dynamical quark mass
used in a numerical simulation. They thus are suitable observables to set 
a scale for other physical observables which may be sensitive to these
external parameters of QCD.

\subsection{Quark mass dependence of the transition temperature}      

\begin{figure}
\begin{center}
\epsfig{file=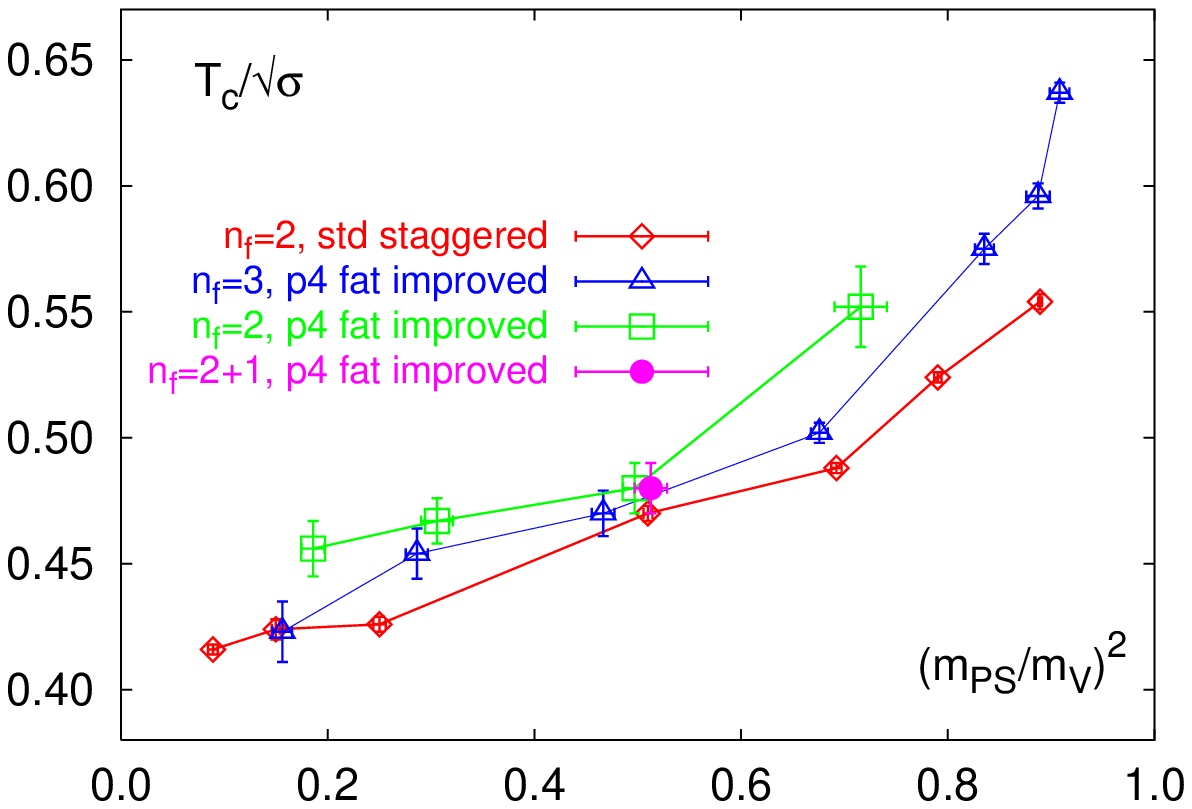,width=74mm}
\epsfig{file=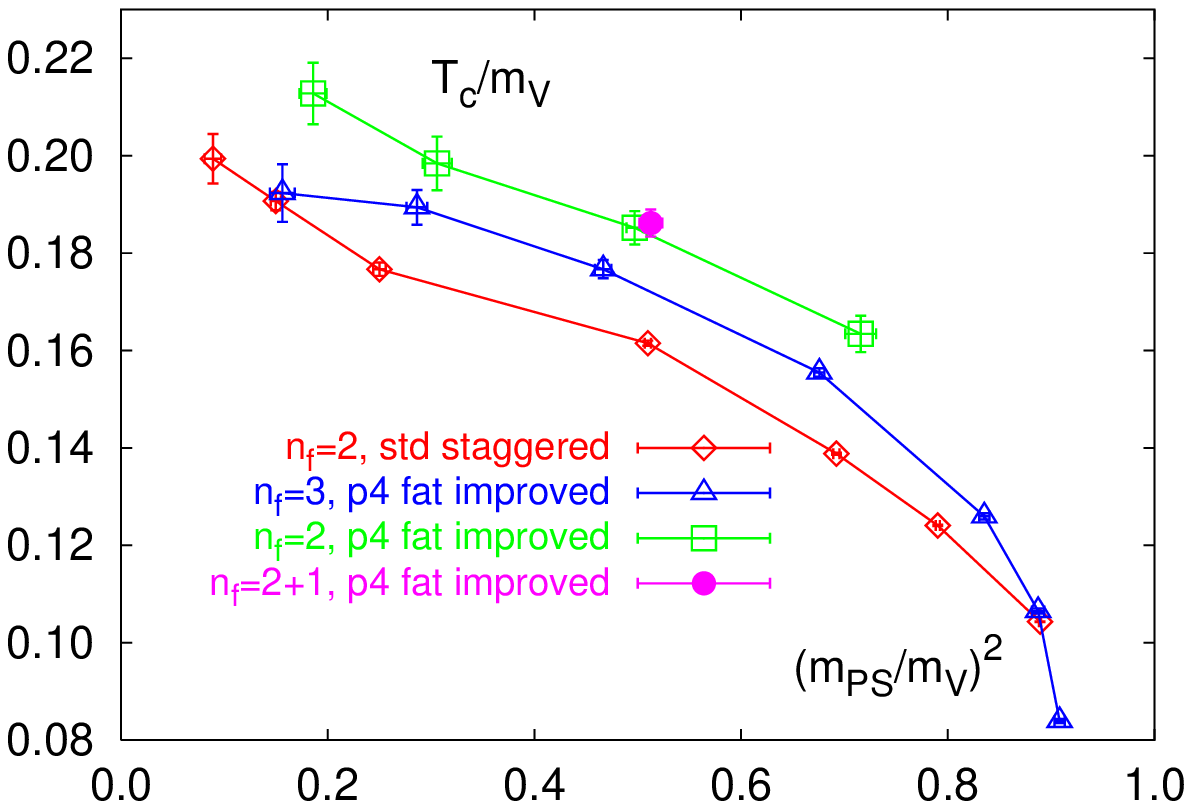, width=74mm}
\end{center}
\caption{
\label{fig:crit_temp}
The transition temperature versus $(m_{\rm PS}/m_{\rm V})^2$ 
for 2 and 3-flavour QCD obtained from calculations with the p4 action
on lattices with temporal extent $N_\tau=4$. Also shown are results 
from calculations using unimproved gauge and staggered fermion 
actions \cite{Luetgemeier1998}.}
\end{figure}

In the previous subsections we have discussed the calculations 
performed to determine the pseudo-critical couplings for the QCD 
transition to the high temperature plasma phase and presented 
zero temperature calculations of meson masses and the string tension
which are needed to set the scale for the transition temperature. 
The resulting
pseudo-critical temperatures expressed in units of the square root of the 
string tension as well as in units of the vector meson mass are given in 
Table~\ref{tab:crit_temp} and shown in Figure~\ref{fig:crit_temp}.
We note that $T_c/\sqrt{\sigma}$ and $T_c/m_{\rm V}$ show a consistent 
flavour dependence. The transition temperature decreases slightly when
the number of flavours is increased. The effect seems to be somewhat 
stronger in the latter ratio, which reflects the 
relative flavour dependence of the two observables used to set
the physical scale (see Figure~\ref{fig:sigma_rho}). The quark mass
dependence or, correspondingly, the dependence on the meson mass ratio 
$m_{\rm PS}/m_{\rm V}$, is, however, completely different for 
$T_c/\sqrt{\sigma}$ and $T_c/m_{\rm V}$. 
In fact, the decrease of $T_c/m_{\rm V}$ with increasing quark mass
mainly reflects the
strong quark mass dependence of $m_{\rm V}$ itself and does not 
represent the physical quark mass dependence of the transition
temperature. The ratio $T_c/\sqrt{\sigma}$, on the other hand, 
shows the expected quark mass 
dependence; with decreasing quark mass the transition temperature drops
from the pure gauge value, $(T_c/\sqrt{\sigma})_{\rm SU(3)}\simeq 0.635$
to a value  $(T_c/\sqrt{\sigma})_{\rm chiral} \simeq (0.4-0.45)$. This 
is in accord with the physical picture that with decreasing quark mass
hadronic degrees of freedom become lighter and thus get more easily  
excited in a thermal heat bath. They then can contribute to the overall energy 
and particle density of the system and thus can trigger the onset of a phase 
transition already at a lower
temperature. Figure~\ref{fig:crit_temp} thus underlines the importance
of a physical observable for setting the scale of $T_c$ which itself is
quark mass independent or at least only shows a weak dependence on
$m_q$. Only in this way we can hope to make contact with phenomenological
models that discuss the quark mass dependence of $T_c$.
We have argued in the previous subsection that the string tension and 
partially quenched meson masses fulfill this requirement. We will mainly
use here the former observable to set a physically sensible scale for $T_c$.
Of course, an analysis of $T_c$ in units of a hadron masses is still of
particular importance in the chiral limit as this provides an experimentally 
well determined input whereas $\sqrt{\sigma}$ is only indirectly accessible 
through heavy quark phenomenology.   

In Figure~\ref{fig:crit_temp} we also have shown results for
$T_c/\sqrt{\sigma}$ obtained from simulations with unimproved gauge
and staggered fermion actions on lattices with temporal extent
$N_\tau=4$. They differ systematically from the results obtained from 
simulations with improved actions, which indicates that we have to
consider the influence of finite cut-off effects on our results. 
In the case of unimproved staggered fermions simulations of 2-flavour 
QCD have also been performed on lattices with larger temporal extent, 
{\it i.e.} at smaller lattice spacing \cite{Ber96}. 
These calculations indeed lead to larger values for $T_c/\sqrt{\sigma}$ 
which are consistent with our results obtained with improved fermions. 
It thus seems that the simulations with improved gauge and staggered
fermion actions lead to a sizeable reduction of cut-off effects. 

We have good control over the transition temperatures down to 
meson mass ratios $m_{PS}/m_V \simeq 0.4$, which corresponds to pion
masses of about 370~MeV when one uses the string tension to set a 
physical scale\footnote{Whenever we use the string tension to convert
to physical units we use $\sqrt{\sigma}= 425$~MeV, which is consistent
with the quenched and partially quenched results given in Eq. \ref{phscale}
and corresponds to $\sqrt{\sigma}/m_\rho = 0.55$.}.   
The results obtained for $T_c/\sqrt{\sigma}$ in 2 and 3-flavour QCD
indicate that the transition temperature is lower in the latter case.
From a linear interpolation of our data at intermediate values of the
quark masses we find
\begin{equation}
\hspace*{-0.6cm}
\Delta(T_c/ \sqrt{\sigma}) \equiv
\biggl( {T_c \over \sqrt{\sigma}} \biggr)_{n_f=2}  -
\biggl( {T_c \over \sqrt{\sigma}} \biggr)_{n_f=3}  \simeq 0.03
\quad {\rm for} \quad 0.4 \le m_{PS}/m_V \le 0.70~
\label{tcdiff}
\end{equation}
Similarly we find $\Delta(T_c/m_V) \simeq 0.015$.  
This corresponds to a temperature difference of approximately $(10-15)$~MeV.

\subsection{Transition temperature in the chiral limit}

\begin{figure}
\begin{center}
\epsfig{file=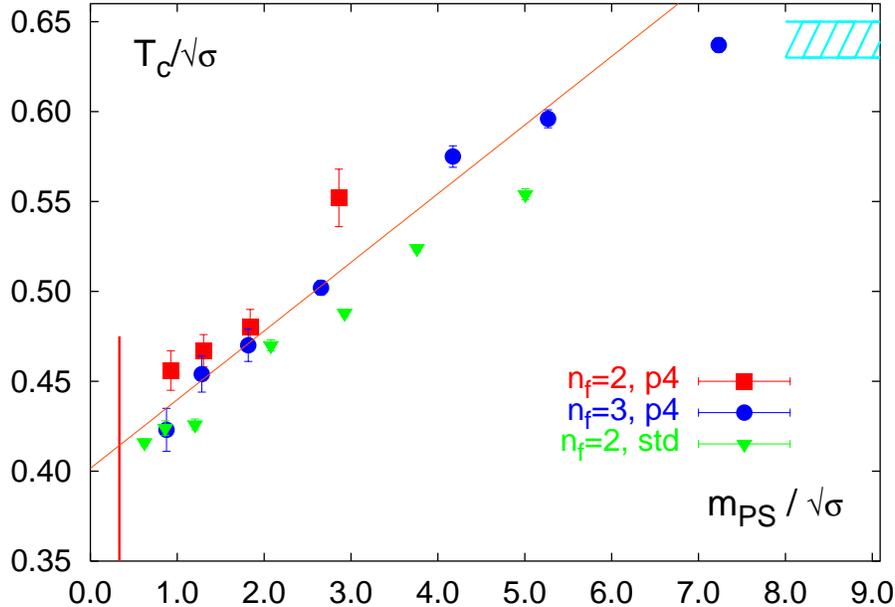,width=125mm}
\end{center}
\caption{
\label{fig:Tc_mpi}
The transition temperature for 2 and 3 flavour QCD in units of the 
string tension versus $m_{\rm PS}/\sqrt{\sigma}$ obtained with standard (std)
\cite{Luetgemeier1998} and improved (p4) staggered
fermions on lattices with temporal extent $N_\tau= 4$.
The hatched band to the right of the figure denotes the quenched
result, the vertical line to the left is the physical
$m_{PS}/\sqrt{\sigma}$ value.}
\end{figure}

As pointed out in section 3.1 the pseudo-critical 
couplings will scale differently close to the chiral limit of  
2 and 3-flavour QCD. This,
of course, translates into a different scaling of the critical
temperatures as a function of the mass of the Goldstone particles,
\begin{equation}
T_c (m_\pi) -T_c(0) \sim \cases{ m_q^{1/\beta\delta} \sim 
m_\pi^{2/\beta\delta} &, $n_f=2$ \cr
m_q \sim  m_\pi^2 &, $n_f \ge 3$}\quad .
\label{tcscaling}
\end{equation} 
However, from Figure~\ref{fig:crit_temp} it is apparent that in 
the mass regime 
currently explored by our simulations the dependence of $T_c$ on the Goldstone 
mass is quite similar for 2 and 3-flavour QCD. As also observed in the
analysis of the pseudo-critical couplings discussed in section 3.1 subleading
corrections will be of particular importance for the chiral extrapolation
in 2-flavour QCD. Subleading contributions linear in $m_q$ will become 
relevant and will influence the dependence of $T_c$ as well as that of the 
pseudo-scalar mass on the quark masses. 
In fact, we find that in a large interval of pseudo-scalar
meson masses $T_c$ depends linearly on $m_{\rm PS}$. This is apparent from 
Figure~\ref{fig:Tc_mpi}. The line shown there is obtained from a fit
to the 3-flavour data in the mass 
interval, $0.85\; <\; m_{\rm PS}/\sqrt{\sigma} \; < \; 5.3$, {\it i.e.}
for $360~{\rm MeV} \; \lsim\; m_{\rm PS} \; \lsim \; 2.3~{\rm GeV}$.  
This gave
\begin{equation}
\biggl( {T_c \over \sqrt{\sigma}} \biggr) (m_{\rm PS}) = 0.40\; (1) +
0.039\; (4) \; \biggl( {m_{\rm PS} \over \sqrt{\sigma}} \biggr)\quad .
\label{tcmpi}
\end{equation}
In view of the quite similar functional dependence of $T_c$ on $m_{\rm
PS}$ for 2 and 3-flavour QCD it seems that this is not indicative for
any form of {\it chiral behaviour}, although an almost linear dependence 
of $T_c$ on $m_{\rm PS}$ is just what one would expect to happen in the 
case of 2-flavour QCD ($2/\beta\delta = 1.1$). We also note the rather weak
dependence of $T_c$ on $m_{\rm PS}$. This suggests that the value of $T_c$ 
is dominantly controlled by a large number of resonances at intermediate
mass values which is not so sensitive to modifications of the mass of the 
lightest states in the spectrum. 
 
\subsubsection{Chiral transition in 3-flavour QCD} 

In order to get some control over systematic errors resulting 
from subleading quark mass dependences in the various observables
used in our analysis we have performed extrapolations to the chiral 
limit in different ways:
\begin{itemize}
\item[i)]{We extrapolate $T_c/\sqrt{\sigma}$ using linear fits 
in $(m_{PS}/m_V)^2$,
$(m_{PS}/\sqrt{\sigma})^2$ and $m_q$ for $m_{PS}/m_V < 0.7$. This 
yields $T_c/\sqrt{\sigma} =0.407 (15)$, 0.419(15) and 0.418(16),
respectively.} 
\item[ii)]{We repeat (i) for $T_c/m_V$. This yields
$T_c/m_\rho = 0.204(5)$, 0.200(3) and 0.200(3).} 
\item[iii)]{We use the vector meson masses calculated
at $\beta_c(m_q=0.05)$ and $\beta_c(m_q=0.025)$ and extrapolate
them to the critical point $\beta_c(m_q\rightarrow 0)=3.258(4)$. 
This yields $T_c/m_\rho = 0.199(4)$.}
\end{itemize}
We note that (i) yields results consistent with the fit linear in
$m_{\rm PS}$ given in Eq.\ref{tcmpi}. 
We also stress that the results obtained from (i) and (ii) imply 
$\sqrt{\sigma}/m_\rho = 0.49(2)$ for the 
ratio of the two zero temperature observables used to set the scale. This 
is consistent with the results shown in Figure~\ref{fig:sigma_rho}.
The above extrapolations also are consistent with the linear
interpolation given in Eq.~\ref{tcmpi} which is shown in 
Figure~\ref{fig:Tc_mpi}.

Combining the results of the various fits performed we estimate for the
critical temperature of 3-flavour QCD in the chiral limit
\begin{equation}
{T_c \over m_\rho} = 0.20 \pm 0.01 \quad \leftrightarrow \quad
T_c = (154 \pm 8)~{\rm MeV}\quad .
\end{equation}
We note that this result has been obtained on lattices with temporal
extent $N_\tau=4$. Although we expect that cut-off effects are small
in our calculation with improved gauge and fermion actions  
a systematic analysis on lattices with larger temporal extent will
be needed to perform a controlled extrapolation to the continuum limit.
However, we estimate that the systematic error resulting from remaining 
cut-off effects is of the same size as the statistical error quoted
above.

\subsubsection{Chiral transition in 2-flavour QCD} 

As indicated in Eq.~\ref{tcscaling}, close to the chiral limit 
we expect to find a universal scaling 
behaviour of $T_c$ that is characteristic for the
universality class of the second order phase transition. For the 
quark mass or pion mass values accessible in present computer
simulations subleading corrections will, however, still be important.
Moreover, the explicit quark mass dependence of the zero temperature
observable used to set the scale has to be taken into account.
This is most drastically seen in the ratio $T_c/m_V$, which 
decreases with increasing pion mass although one clearly expects
that the critical temperature in physical units increases with
increasing mass of the lightest particles in the heat bath. This 
is correctly reflected in the ratio $T_c/\sqrt{\sigma}$. 
In the entire quark mass regime analyzed so far the quark mass
dependence of the ratio $T_c/m_V$, however, is dominated by
the explicit quark mass dependence of $m_V$. 
An appropriate ansatz for the quark mass dependence of $T_c/m_V$ is 
\begin{equation}
\biggl( {T_c\over m_V}\biggr)(m_q) = \biggl( {T_c\over m_V}\biggr)(0)
+ a\; m_q^{1/\beta\delta} + b\; m_q \quad ,
\end{equation}
where one should find $a>0$ and $b<0$.  Thus, also this ratio will 
ultimately approach the chiral limit from above. With the presently 
used quark masses, however, the subleading term, linear in $m_q$, apparently 
still gives the dominant contribution. For 2-flavour QCD the ratio $T_c/m_V$ 
thus clearly does not yet reflect the expected leading order quark mass 
dependence of $T_c$. 

In performing the extrapolation to the chiral limit we followed
the strategy outlined in the previous section. For the case of the
ratio $T_c/\sqrt{\sigma}$ we have performed extrapolations linear
in $m_q$ as well as assuming the $O(4)$ universal behaviour $\sim
m_q^{0.55}$. In the case of $T_c/m_V$, however, we only use
the linear extrapolation. From this we find
\begin{equation}
{T_c \over m_\rho} = 0.225\pm 0.010 \quad , \quad 
{T_c \over \sqrt{\sigma}} = 0.425 \pm 0.015
\end{equation}
We note again that this implies for the ratio of the two observables
used to set the scale, $\sqrt{\sigma}/m_\rho = 0.53(3)$. Using the rho
meson mass to convert the critical temperature to physical
units we find
\begin{equation}
T_c = (173 \pm 8)~{\rm MeV} \quad .
\end{equation}
This result is consistent with earlier estimates obtained from
calculations with the standard staggered fermion action \cite{Kar99}
and also with a recent calculation performed with clover improved
Wilson fermions and a renormalization group improved gauge action \cite{Ejiri}.
We stress again that the errors quoted above are statistical only. Similar to 
the $n_f=3$ case we expect that systematic errors are of similar size.

Let us finally comment on the critical temperature in the physically most
interesting case of (2+1)-flavour QCD. The simulations performed in this
case still use too heavy quarks in the light quark sector, {\it i.e.} $m_{u,d}=0.1$. 
However, even in this case we find that the pseudo-critical temperature is closer to
that of 2-flavour QCD with the same quark mass than to the 3-flavour case.
We thus expect that the critical temperature of 2-flavour QCD provides a good
approximation to the physically realized case of two light and a heavier
strange quark.  

\section{The heavy quark free energy}

We have discussed in the previous sections the influence of 
dynamical quarks on the QCD transition temperature. In addition
to the quark mass dependence of $T_c$ we also expect that the
presence of dynamical quarks of finite mass in the thermodynamic 
heat bath will lead to    
large qualitative and quantitative modifications of
the heavy quark potential\footnote{We stick here to the commonly used
notion of a heavy quark potential, although at non-zero temperature we 
are actually dealing with the heavy quark free energy \cite{McL81}.} $V(r,T)$. 
In the pure gauge ($m_q\rightarrow \infty$) limit the heavy quark potential 
is strictly confining for all temperatures below $T_c$. For any finite
quark mass value, however, string breaking will occur 
and the potential will approach a constant value for
$r \rightarrow \infty$. 

\begin{figure}
\begin{center}
\epsfig{file=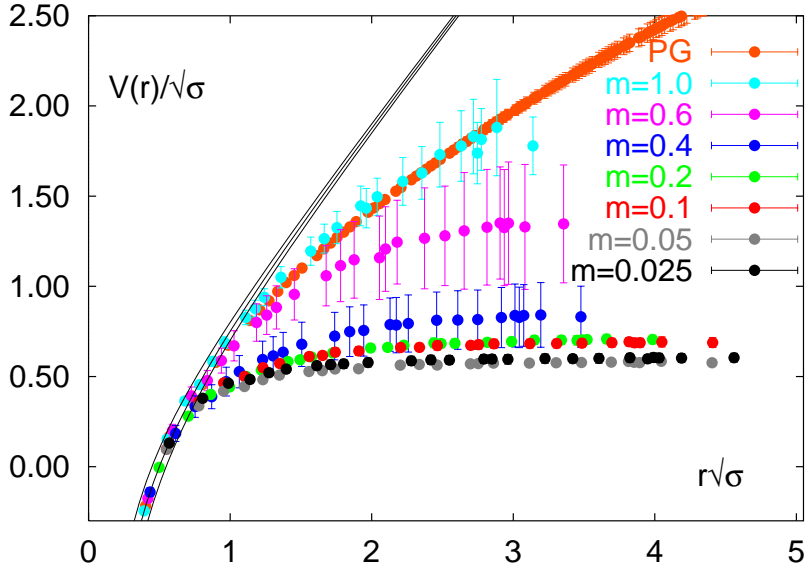,width=115mm}
\end{center}
\caption{
\label{fig:potential_nf3_m}
Quark mass dependence of the heavy quark potential for three
flavour QCD below the transition temperature at a temperature
$T\simeq 0.97~T_c$. The band of lines gives the
Cornell-potential in units of the square root of the string tension,
$V(r)/\sqrt{\sigma} = -\alpha / r\sqrt{\sigma} + r\sqrt{\sigma}$ with
$\alpha = 0.25\pm 0.05$.The gauge couplings corresponding to the
different quark mass values are $\beta =$~4.06, 3.97, 3.86, 3.76, 3.59,
3.46, 3.38, 3.32 (from top to bottom).}
\end{figure}

While it seems to be difficult to detect string breaking in calculations 
of the heavy quark potential at zero temperature using Wilson loops, 
string breaking does quite naturally occur at finite temperature, where 
the heavy quark potential can be extracted from Polyakov loop correlation 
functions
\begin{equation}
{V_{q\bar{q}} (r,T) \over T} = - \ln \langle
L(\vec{x})L^{\dagger}(\vec{y})
\rangle + {\rm const.}~~,~~{\rm with~~} |\vec{x} - \vec{y}|=r~~.
\label{polpot}
\end{equation}
In fact, string breaking is directly related to the non-vanishing
of the Polyakov loop expectation value at any temperature and
finite quark mass,
\begin{equation}
\lim_{r \rightarrow \infty} \biggl( {V_{q\bar{q}} (r,T) \over T} \biggr)
= -2 \ln | \langle L \rangle | + {\rm const.} \quad .
\label{asypolpot}
\end{equation}
First results on string breaking at finite temperature have been reported 
in \cite{DeT99}.

\begin{figure}
\epsfig{file=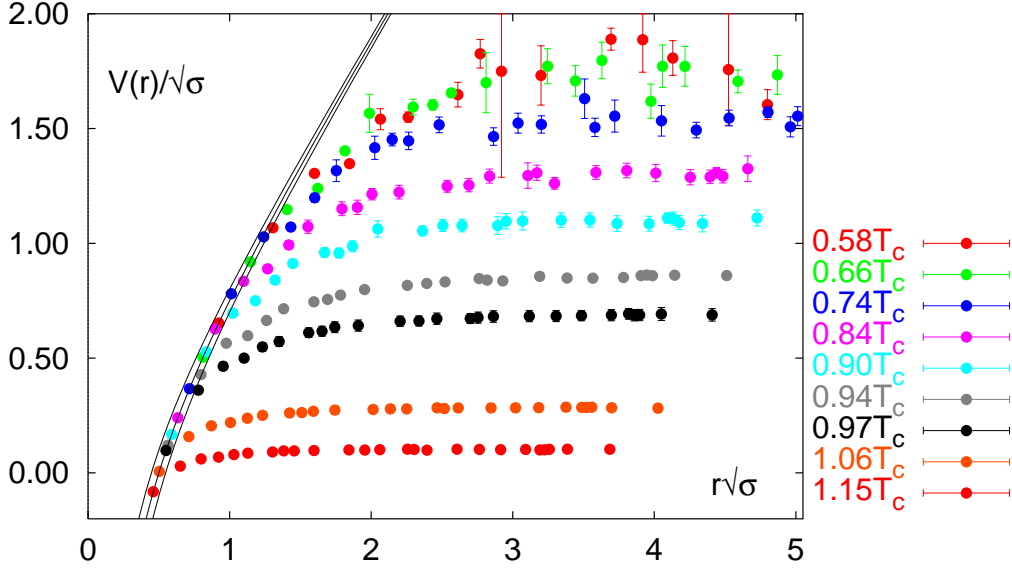,width=115mm}
\caption{
\label{fig:potential_nf3_T}
Temperature dependence of the heavy quark potential for three
flavour QCD with a quark mass $m_q=0.1$. The band of lines gives the
Cornell-potential in units of the square root of the string tension,
$V(r)/\sqrt{\sigma} = -\alpha / r\sqrt{\sigma} + r\sqrt{\sigma}$ with
$\alpha = 0.25\pm 0.05$. The gauge couplings corresponding to the
different temperatures are $\beta =$~3.25, 3.30, 3.35, 3.40, 3.43,
3.45, 3.46, 3.50, 3.54. The string tension values used to set the scale
are based on the interpolation formula given in Ref. 12. The 
potentials have been normalized at short distances such that they
agree with the zero temperature Cornell potential at $r=1/4T$.}
\end{figure}

In order to analyze the quark mass dependence of the heavy quark
potential and its structure at large distances we have calculated 
Polyakov-loop correlation functions for 3-flavour QCD
at a temperature close but below the transition temperature, 
$T \simeq 0.97~T_c$. This analysis has been performed with the p4-action
on lattices of size $16^3\times 4$ as a by-product of our determination
of the critical couplings at various quark mass values. In addition we used
the string tensions obtained from our zero temperature calculations
at a nearby value of the gauge coupling, corresponding to $T=T_c$,
to set the scale for the potential and the spatial separation $r$
of the static quark anti-quark pair. These potentials are shown in
Figure~\ref{fig:potential_nf3_m}. They have been normalized such that
they agree with the zero temperature Cornell potential at short
distances, {\it i.e.} at $r=1/ 4 T$.

We note that with decreasing quark mass the screening effects
resulting from the presence of virtual quark anti-quark pairs
lead to a more rapid flattening of the potential. While for
$m_q=1.0$ the potential still coincides with the finite temperature
potential in the pure gauge sector \cite{Kac00} at least up to
$r\sqrt{\sigma} =3$, the influence of dynamical quarks becomes large 
already for $m_q=0.6$. At small quark masses
the screened potential reaches a limiting form 
and starts deviating from the pure gauge heavy quark potential
already at $r\sqrt{\sigma} \simeq 0.7$ and saturates at
$r\sqrt{\sigma} \simeq 1.5$. This corresponds to distances $r \simeq
0.3$~fm and 0.7~fm, respectively. In general we find that deviations
from the chiral limit are small for quark masses below
$m_q  \simeq 0.1$ or $m_{\rm PS} \simeq 1.8 \sqrt{\sigma}$.

The above analysis of the quark mass dependence has been performed at a
fixed temperature close to $T_c$. In addition screening effects are, of
course, also strongly temperature dependent. In the pure gauge
case this leads to a temperature dependence of the string tension
\cite{Kac00}. In the presence of dynamical quarks we find that 
string breaking sets in at larger distances when the temperature is
lowered. Note that this happens although at fixed bare quark mass
$m_q$ the physical quark mass decreases with decreasing temperature. 
Thermal effects seem to become small for temperatures
below $T \simeq 0.65~T_c$. This becomes evident from
Figure~\ref{fig:potential_nf3_T} where we show results from a
calculation for 3-flavour QCD with quark mass $m_q=0.1$. In the low
temperature regime the potential starts deviating from simple
Cornell-type confinement potential for $r\sqrt{\sigma} \simeq 1.2$
and reaches a plateau for $r\sqrt{\sigma} \simeq 3$. This corresponds to
distances $r \simeq 0.6$~fm and 1.4~fm, respectively.

\begin{figure}
\epsfig{file=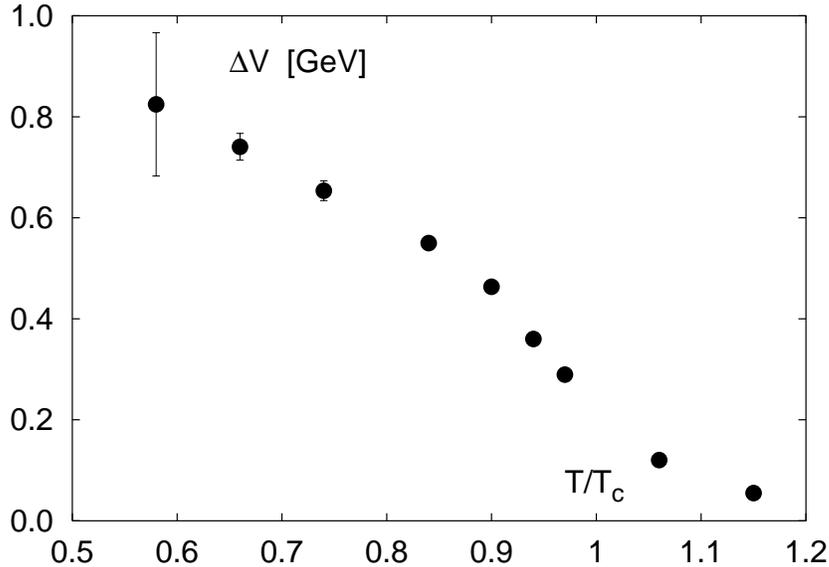,width=115mm}
\caption{
\label{fig:diss_nf3_T}
Temperature dependence of the depth of the heavy quark potential,
defined in Eq.~\ref{deltaV},
calculated in 3-flavour QCD with quarks of mass $m_q=0.1$. In order
to convert to physical units we have used $\sqrt{\sigma}= 425$~GeV
as input.}
\end{figure}

To quantify the screening effect in the presence of light dynamical
quarks we define the depth of the potential as the difference between
the asymptotic value at large distances and the 
potential at distance $r\sqrt{\sigma}=0.5$, 
\begin{equation}
\Delta V \equiv \lim_{r\rightarrow \infty}V(r) -V(0.5/\sqrt{\sigma})
\label{deltaV}
\end{equation}
which is the point at which the Cornell-potential vanishes if one
chooses the coupling of the Coulomb term as $\alpha = 0.25$. It 
approximately corresponds to a distance $r\simeq 0.23$~fm. 
In Figure~\ref{fig:diss_nf3_T} we show this difference of 
potential energies. 
We note that $\Delta V$ changes most rapidly in the vicinity of $T_c$.
At $T_c$ we obtain
$\Delta V \simeq 0.5\sqrt{\sigma} \simeq 200~{\rm MeV}$ which, in fact,
is compatible with the depth of a simple unscreened Coulomb
potential, $V(r)/\sqrt{\sigma} = -\alpha / r\sqrt{\sigma}$ with 
$\alpha = 0.25$.

\section{Conclusions}

We have analyzed the quark mass and flavour dependence of the critical
temperature of QCD using improved gauge and staggered fermion actions.
We find that in a large interval of quark mass values, which corresponds
to zero-temperature pseudo-scalar meson masses $m_{\rm PS}$ between 
360~MeV and about 2.3~GeV 
the transition temperature $T_c(m_{\rm PS})$ depends only weakly on
$m_{\rm PS}$. In this entire mass interval the transition temperature
in two-flavour QCD is about 10\% larger than that of three-flavour QCD.
An extrapolation of both pseudo-critical temperatures to the chiral 
limit yields $T_c(0) = (173\pm 8)$~MeV in the case of two-flavour QCD
and $T_c(0) = (154\pm 8)$~MeV for three-flavour QCD. In both cases we 
have quoted statistical errors only. From a comparison with calculations
performed with unimproved actions and the experience gained with
discretization errors in that case as well as in the pure gauge sector
we estimate that systematic errors inherent in the present analysis are
of the same order of magnitude. The results obtained here with
staggered fermions are also consistent with results obtained from 
studies of two-flavour QCD with clover-improved Wilson fermions
\cite{Ejiri}.

Our analysis of the temperature and quark mass dependence of the heavy 
quark free energy shows that screening at $T\simeq 0.97 T_c$ is very 
efficient already in a quark mass regime where the (zero-temperature) 
pseudo-scalar meson
masses are about 1~GeV. Furthermore, screening effects become 
approximately quark mass independent for smaller masses. 
For a fixed value of the bare quark mass, $m_q=0.1$, we find that
string breaking effects set in at a distance of about 0.6~fm for  
small temperatures, $T\; \lsim \; 0.6T_c$. With increasing temperature
the onset point gradually shifts to shorter distances. At $T_c$ the heavy 
quark free energy gets screened already at distances of about 0.3~fm.

\vspace{0.5cm}
\noindent
{\bf Acknowledgements:}

\medskip
\noindent
The work has been supported by the TMR network
ERBFMRX-CT-970122 and by the DFG under grant Ka 1198/4-1. 

\vspace{0.7cm}

\end{document}